\documentclass{PoS}

\title{Re-analysis of Timing Parameters of OAO 1657-415}

\ShortTitle{Re-analysis of Timing Parameters of OAO 1657-415}

\author{\speaker{A. Baykal}\\
        Physics Department, Middle East Technical University, Ankara, Turkey\\
        E-mail: \email{altan@astroa.physics.metu.edu.tr}}

\author{S.\c{C}. \.{I}nam\\
        Department of Electrical and Electronics Engineering, Ba\c{s}kent University, Ankara, Turkey\\
        E-mail: \email{inam@baskent.edu.tr}}

\author{B. \.{I}\c{c}dem\\
        Physics Department, Middle East Technical University, Ankara, Turkey}
        
\author{E. Beklen\\
	    Physics Department, S\"{u}leyman Demirel University, Isparta, Turkey}
	    
\abstract{In this paper, we present the re-analysis of a set of RXTE-PCA observations of OAO 1657-415 during 1997 August 20-27. We found a revised timing solution which was found to fit better to the data and updated pulse frequency values accordingly. We also verified that there is a marginal correlation between the gradual spin-up (or decrease in spin-down rate) and increase in X-ray luminosity as suggested by Baykal (2000).}

\FullConference{Fast X-ray timing and spectroscopy at extreme count rates: Science with the HTRS on the International X-ray Observatory - HTRS2011,\\
		February 7-11, 2011\\
		Champ\'ery, Switzerland}

\begin{document}
\section{Introduction}

The accretion powered pulsar OAO 1657-415 was discovered with the Copernicus satellite (Polidan et al. 1978). Its spin period was found to be 38.22s (White \& Pravdo et al. 1979). Orbital parameters and the eclipsing nature of the source was first reported by Chakrabarty et al. (1993). Bildsten et al. (1997) used BATSE observations to improve the orbital parameters of the source and found the orbital period to be $P_{orb}=(10.44809\pm0.00030)$days.

The companion of the source was found to be a highly reddened B supergiant (Chakrabarty et al. 2002). From ASCA observations, Audley et al. (2006) found a dust-scattered X-ray halo which led them to estimate the distance to the source as $7.1\pm 1.3$ kpc.  

Baykal (1997) suggested an explanation of spin-up and spin-down episodes in the BATSE data as the formation of episodic accretion disks while the source accretes from its companion's stellar wind. Using BATSE observations, Inam \& Baykal (2000) showed that the specific angular momentum of the accreted material is correlated with the torque exerted on the neutron star which is also interpreted as a sign of episodic accretion disks. Barnstedt et al. (2008) found that the long-term period evolution of OAO 1657-415 is actually characterised by a long-term spin-up with a rate of $\dot{P}_{mean}=-1.24\times10^{-9}{\rm{s s}}^{-1}$ overlayed by sets of relative spin-down/spin-up episodes, which appear to repeat quasi-periodically on a 4.8 yr time scale. 

Baykal (2000) analyzed RXTE-PCA observations of OAO 1657-415 during 1997 August 20-27. Analysis of these observations revealed that the source was in an extended phase of spin-down. From the marginal correlation between the gradual spin-up (or a decrease in spin-down rate) and increase in X-ray luminosity, it was inferred by Baykal (2000) that OAO 1657-415 was observed during a stable accretion episode in which the prograde accretion disk was formed. 

In this paper, we present a revised timing solution and the results of corresponding analysis of the RXTE-PCA observations that were studied before by Baykal (2000). In the next section, we present observations and timing analysis. In Section 3, we discuss our results.

\begin{table}[htb]
\caption{Timing Solution of OAO 1657-415 for RXTE Observations$^{a}$}
\label{ts}
\[
\begin{tabular}{c|c|c}  \hline 
Parameter & Previous Work$^b$ & This Work \\ 
\hline
Orbital Epoch (MJD) & 48515.99(5)$^{b}$ &  50689.116(50) \\
P$_{orb}$ (days)    & 10.44809(30)$^{b}$ & 10.44749(55) \\
a$_{x}$ sin i (lt-sec) & 106.0(5)$^{b}$  & 106.10(2) \\
e                       & 0.104(5)$^{b}$  & 0.1033(6) \\
w   & 93(5)$^{b}$ & 87.6(1.3) \\ 
Epoch(MJD)  & 50683.95400(2) & 50680.322(6) \\
Pulse Frequency (Hz) & 0.026775618(4) & 0.026777129(9) \\
Pulse Freq. Derivative (Hz s$^{-1}$)  & -3.27(9)$\times 10^{-12}$ & -3.21(7)$\times 10^{-12}$ \\
Reduced $\chi^2$ & 24 & 5.53 \\ 
\hline \hline
\end{tabular}
\]
$^{a}$ Confidence intervals are quoted at the 1 $\sigma $ level. \\
$^{b}$ This timing solution was obtained by Baykal (2000). \\
$^{c}$ Orbital parameters are taken from Bildsten et al. (1997).\\
P$_{orb}$=orbital period, a$_{x}$ sin i=projected semimajor axis,
e=eccentricity, w=longitude of periastron.\\
\end{table}

\section{Observation and Analysis}

\begin{figure}[hbt]
\label{pat}
\begin{tabular}{cc}
\hspace{-0.4cm}\includegraphics[width=6.4cm,keepaspectratio=true,angle=0]{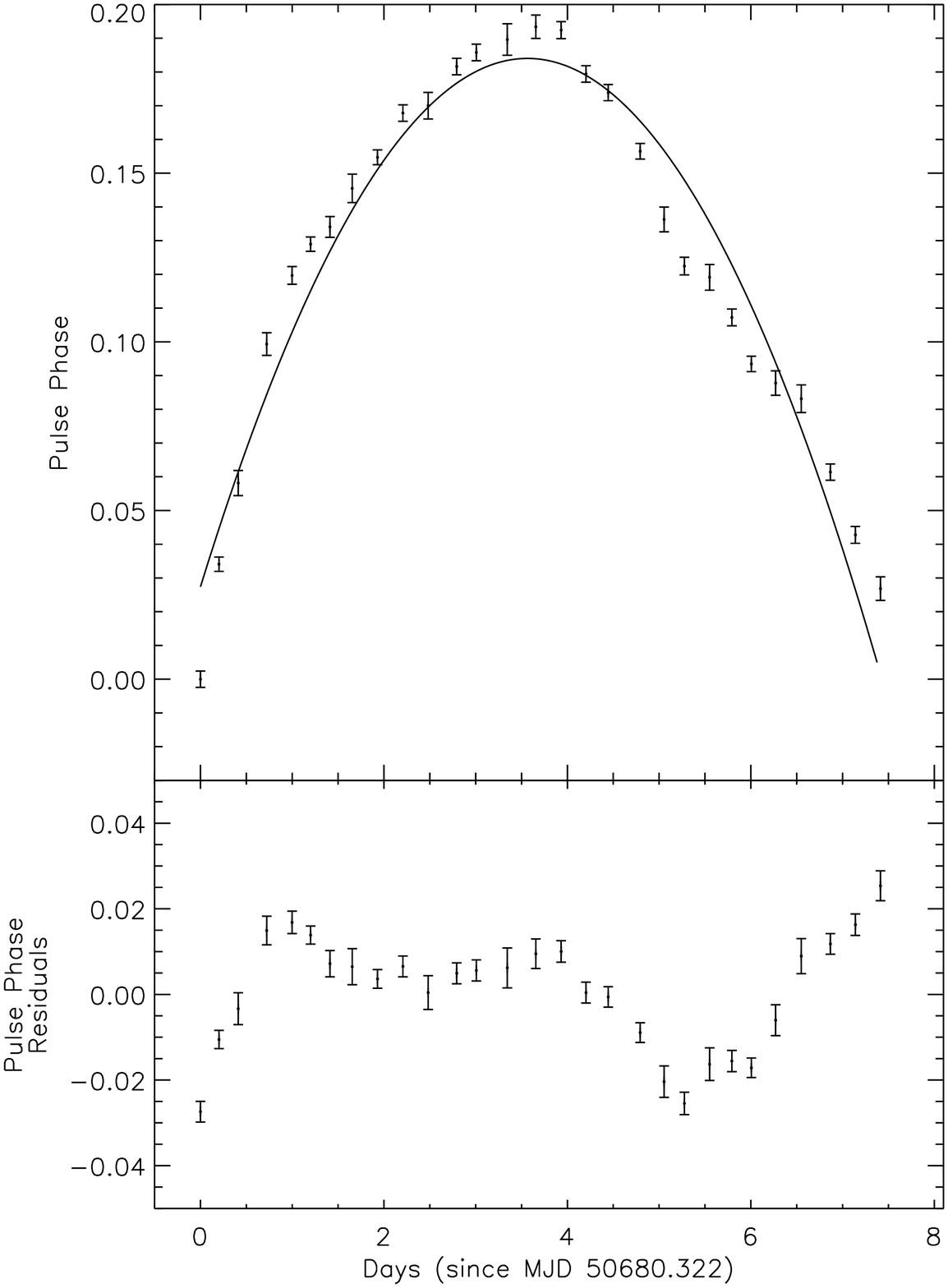} &
\hspace{-0.4cm}\includegraphics[width=6.4cm,keepaspectratio=true,angle=0]{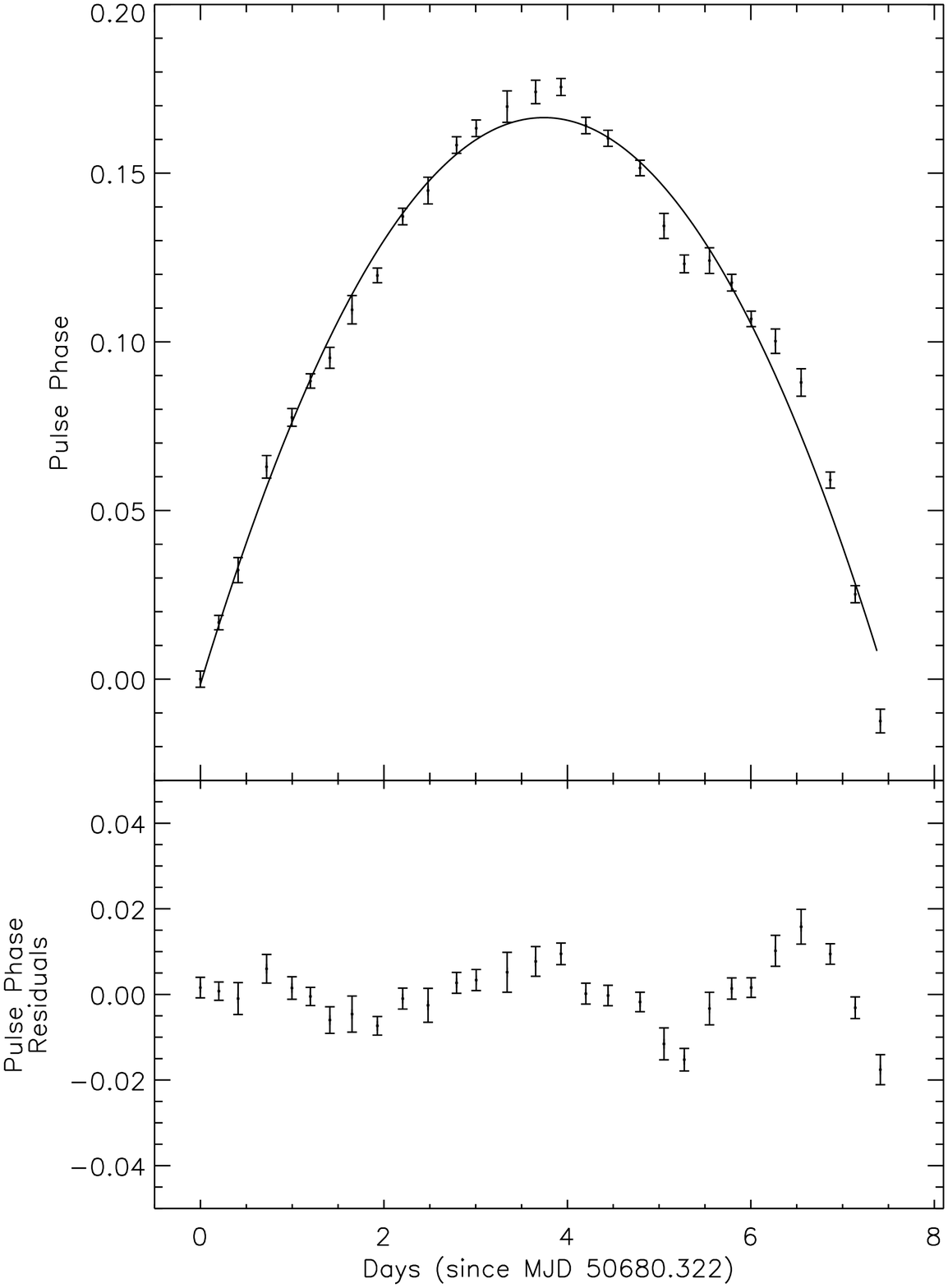} \\
\includegraphics[height=6cm,keepaspectratio=true,angle=-90]{old_freq.ps} &
\includegraphics[height=6cm,keepaspectratio=true,angle=-90]{new_freq.ps}
\end{tabular}\\
\caption{{\bf{(top)}} Phase offsets in pulse arrival times, {\bf{(bottom)}} and pulse frequency measurements of OAO 1657-415. Plots on the left are taken from Baykal (2000) and plots on the right show the results of the current work. Solid line denotes the best fit of the arrival times.}
\end{figure}

The results presented here are based on data collected with the Proportional Counter Array (PCA, Jahoda et al. 1996) between 1997 August 20-27 which were analyzed before by Baykal (2000). The lightcurves were background subtracted and the time columns of the lightcurves are corrected to the barycenter of the solar system. To find the pulse arrival times, we use the
method that makes use of harmonic representation of pulse profiles (Deeter \& Boynton, 1985). In this method, pulse profiles are expressed in terms of harmonic series and cross-correlated with the master pulse profile to obtain pulse arrival times. Using this method we obtained 29 pulse arrival times (one for each RXTE orbit). 

To subtract the effect of orbital motion to the pulse arrival times we make use of the expression used by Deeter et al. (1981). Using this expression, we revise the timing solution of the source and find that the new timing solution in Table \ref{ts} gives a better fit, compared to the timing solution of Baykal (2000), to the pulse arrival times that were presented in Figure \ref{pat}. The revised spin-down rate during the observations is found to be $\dot{\nu}=-(3.21\pm0.07)\times10^{-12}{\rm{Hz s}}^{-1}$. 

For every 4 or 5 consecutive pulse arrival times, we estimate a pulse frequency value (see Figure \ref{pat}). It is important to note that these pulse frequency values exactly correspond to the X-ray flux values found before (see Figure 5 in Baykal 2000). 

\section{Discussion}

In this paper, we present the re-analysis of a set of RXTE-PCA observations of OAO 1657-415 during 1997 August 20-27. The new timing solution is found to fit better to the data (see Table \ref{ts}). Resulting pulse frequencies on the right shown in Figure \ref{pat} show a similar time dependence compared with the pulse frequencies obtained using the old timing solution on the left. Thus,  
using the new timing solution, we can still conclude with the help of Figure 5 in Baykal (2000) that there is a marginal correlation between the gradual spin-up (or decrease in spin-down rate) and increase in X-ray luminosity.

\begin{acknowledgments}
  We acknowledge research project TBAG 109T748 of the Scientific and
Technological Research Council of Turkey (T\"{U}B\.{I}TAK).
\end{acknowledgments}

\end{document}